\documentclass[amsmath,amssymb,superscriptaddress,aps,prx,reprint,floatfix]{revtex4-2}

\usepackage{graphicx}
\usepackage{dcolumn}
\usepackage{glossaries}
\usepackage[version=3]{mhchem}
\usepackage[per-mode=symbol]{siunitx}
\usepackage[most]{tcolorbox}
\usepackage{xcolor}
\usepackage{hyperref}
\usepackage{booktabs}

\graphicspath{{submission-v1/figures}}

\hypersetup{
    unicode,
    colorlinks=true,
    urlcolor=black,
    linkcolor=black,
    citecolor=black
}

\newcommand{\bzsse}[0]{\ce{BaZrS_{3$x$}Se_{3{--}3$x$}}}

\newacronym{acf}{ACF}{autocorrelation function}
\newacronym{dft}{DFT}{density-functional theory}
\newacronym{haadf}{HAADF}{high-angle annular dark-field}
\newacronym{mbe}{MBE}{molecular beam epitaxy}
\newacronym{mc}{MC}{Monte Carlo}
\newacronym{md}{MD}{molecular dynamics}
\newacronym{mcmd}{MCMD}{Monte Carlo molecular dynamics}
\newacronym{mlip}{MLIP}{machine-learned interatomic potential}
\newacronym{nep}{NEP}{neuroevolution potential}
\newacronym{sgc}{SGC}{semi-grand canonical}
\newacronym{soc}{SOC}{spin-orbit coupling}
\newacronym{sro}{SRO}{short-range order}
\newacronym{stem}{STEM}{scanning transmission electron microscopy}
\newacronym{ti}{TI}{thermodynamic integration}

\makeatletter
\let\oldtheequation\theequation
\renewcommand\tagform@[1]{\maketag@@@{\ignorespaces#1\unskip\@@italiccorr}}
\renewcommand\theequation{(\oldtheequation)}
\makeatother

\DeclareSIUnit\angstrom{\text{Å}}
\DeclareSIUnit\site{\text{site}}
\DeclareSIUnit\atom{\text{atom}}

\newcommand{\hmn}[1]{
  \ensuremath{\begingroup\setupHMN #1\endgroup}%
}

\newcommand{\setupHMN}{%
  \doHMN{-}{\HMNoverline}%
  \doHMN{*}{\HMNminverse}%
  \doHMN{i}{\infty}
}

\newcommand{\doHMN}[2]{%
  \begingroup\lccode`~=`#1
  \lowercase{\endgroup\let~}#2%
  \mathcode`#1="8000
}

\newcommand{\HMNminverse}[1]{\frac{#1}{m}}
\newcommand{\HMNoverline}[1]{\mkern1mu\overline{\mkern-1mu#1\mkern-1mu}\mkern1mu}

\begin{document}

\title{Anion Ordering and Phase Stability Govern Optical Band Gaps in \texorpdfstring{\bzsse{}}{BaZrS(3x)Se(3-3x)}}

\newcommand{\addchalmers}{Department of Physics and Astronomy, Chalmers University of Technology, SE-41296, Gothenburg, Sweden}
\newcommand{\addnorthumbria}{Department of Engineering, Physics and Mathematics, Northumbria University, Newcastle upon Tyne, NE1 8QH, United Kingdom}
\newcommand{\addmit}{Department of Materials Science and Engineering, Massachusetts Institute of Technology, Cambridge, MA 02139, United States of America}
\newcommand{\adducl}{Department of Chemistry, University College London, London WC1E 6BT, United Kingdom}

\author{Erik Fransson}
\affiliation{\addchalmers}
\author{Michael Xu}
\affiliation{\addmit}
\author{Prakriti Kayastha}
\affiliation{\addnorthumbria}
\affiliation{\adducl}
\author{Kevin Ye}
\author{Ida Sadeghi}
\author{Rafael Jaramillo}
\author{James M. LeBeau}
\affiliation{\addmit}
\author{Lucy Whalley}
\affiliation{\addnorthumbria}
\author{Paul Erhart}
\email{erhart@chalmers.se}
\affiliation{\addchalmers}

\date{\today}

\begin{abstract}
Chalcogenide perovskites have emerged as promising lead-free materials for photovoltaic and thermoelectric applications.
Among them, \ce{BaZrS3} has attracted particular attention due to its thermal and chemical stability, favorable optoelectronic properties, and low thermal conductivity.
Here, we combine molecular dynamics and Monte Carlo simulations based on a machine-learned interatomic potential with scanning transmission electron microscopy to investigate mixing thermodynamics and phase stability in the \bzsse{} system.
We identify an unusual ordered structure that persists at room temperature, most prominently at \qty{33}{\percent} S, where S and Se atoms form alternating layers within the crystal.
Free-energy calculations yield the temperature–composition phase diagram, including a non-perovskite $\delta$ phase in the Se-rich limit and a perovskite phase in the S-rich limit, separated by a broad two-phase region.
Analysis of the dielectric function and the absorption coefficient demonstrates that composition, crystal structure, and anion ordering jointly control the optical band gap.
Selenium alloying enables tuning between approximately \num{1.6} and \qty{1.9}{\electronvolt}, while anion ordering within a given composition reduces the gap by about \qty{0.12}{\electronvolt}.
Lastly, variations between structural polymorphs give rise to band gap differences of up to \qty{0.4}{\electronvolt}.
\end{abstract}

\maketitle

\section{Introduction}

Perovskites with the general formula \ce{ABX3} constitute a versatile class of materials for optoelectronic applications.
Hybrid lead–halide perovskites, in particular, have enabled record-breaking single-junction and tandem photovoltaic efficiencies \cite{nrel-efficiencies}.
However, their long-term chemical instability and the presence of toxic lead remain significant barriers to large-scale deployment.
These limitations have motivated the search for alternative perovskite chemistries that retain favorable optoelectronic properties while improving chemical robustness and sustainability.

Chalcogenide perovskites (X = S, Se) provide one such platform.
These materials combine strong optical absorption with improved environmental stability and are composed of earth-abundant, non-toxic elements \cite{sopiha2022chalcogenide, jaramillo2019praise, niu2018optimal, kayastha2023high, osei2021examining, wu2023ultralow}.
Among them, \ce{BaZrS3} has emerged as a model system, exhibiting strong band-to-band photoluminescence and favorable defect physics \cite{nielsen2025bazrs3, Nishigaki2020, yuan2024assessing, Mualem2025Using}.
However, its band gap of \qty{1.9}{\electronvolt} \cite{sadeghi_making_2021, Agarwal2025} lies above the ideal range for single-junction photovoltaics, as the Shockley–Queisser limit places the maximum theoretical efficiency at \qtyrange{1.3}{1.4}{\electronvolt}.
Band gap engineering through alloying is therefore a natural strategy to enhance its photovoltaic potential.
Partial substitution on the B-site with Ti \cite{wei2020ti, zilevu2024solution} or on the X-site with Se \cite{sadeghi2023expanding} reduces the band gap into the optimal range \cite{Meng2016, Li2022, sadeghi2023expanding, ye2024processing, Agarwal2025}.
An additional benefit is that Se incorporation is predicted to weaken electron–phonon coupling, potentially suppressing non-radiative recombination in \bzsse{} alloys \cite{Li2022}.
While alloying thus improves optoelectronic properties, it also modifies bonding and lattice energetics, raising fundamental questions about phase stability and anion ordering.

The structural stability of the perovskite sulfide end member is well established.
At ambient conditions, \ce{BaZrS3} adopts the corner-sharing perovskite structure with orthorhombic \hmn{Pnma} symmetry \cite{Jess2022}.
Upon heating, it undergoes displacive phase transitions to tetragonal and cubic polymorphs \cite{jaykhedkar2023temperature, bystricky2024thermal, Jaiswal2024High, Kayastha2025}.

In contrast, there are conflicting structures reported for the perovskite selenide end member.
An early experimental attempt to synthesize stoichiometric \ce{BaZrSe3} via a solid-state reaction yielded off-stoichiometric, needle-like structures composed of face-sharing octahedra, rather than the corner-sharing perovskite network \cite{Tranchitella1998}.
More recently, stoichiometric \ce{BaZrSe3} in a closely related face-sharing hexagonal phase has been reported for nanocrystalline samples prepared by hot injection \cite{Agarwal2025Hexagonal}.
First-principles studies constrained to the 1:1:3 stoichiometry instead predict a different edge-sharing polymorph with \hmn{Pnma} symmetry \cite{Ong2019, Liu2023}.
Se-rich corner-sharing perovskite phases, including stoichiometric \ce{BaZrSe3}, have been realized using non-equilibrium methods \cite{sadeghi2023expanding, ye2024processing}.
Although these perovskite phases are likely metastable, they exhibit long-term stability at ambient conditions.
For the mixed-anion series \bzsse{}, powder synthesis studies indicate phase segregation above approximately \qty{40}{\percent} Se \cite{Nishigaki2020}, also implying limited stability of the perovskite framework a high Se content under near-equilibrium conditions. 

Collectively, the available evidence suggests that the perovskite form of \ce{BaZrSe3} is not the equilibrium ground state under ambient conditions, but that it is kinetically stable.
However, a unified thermodynamic description capable of reconciling equilibrium phase segregation, metastable perovskite formation, and possible anion ordering across composition and temperature has not yet been established.
Developing such a description is essential for the optimization of mixed-anion chalcogenide perovskites and their optical properties.

Here, we establish the temperature–composition phase behavior of the mixed \bzsse{} system combining atomic scale modeling and \gls{stem} analysis.
Building on our previously developed dataset for \ce{BaZrS3} \cite{Kayastha2025}, we train a \gls{mlip} with \gls{dft} data.
This approach enables large-scale sampling of the configurational and vibrational degrees of freedom while retaining close to hybrid-functional accuracy.
Using \gls{md} and \gls{mcmd} simulations, we explore anion configurations and competing structural motifs over a broad range of temperatures and compositions, keeping the Ba:Zr:(S+Se) ratio fixed at 1:1:3.
We predict the emergence of an unusual trans-ordered S/Se arrangement at \qty{33}{\percent} S content, which we confirm experimentally by \gls{stem}.
Furthermore, we construct a temperature–composition phase diagram that rationalizes the experimentally observed stability limits of the perovskite phase and clarifies the thermodynamic origins of metastability in Se-rich compositions.
Together, these results provide quantitative thermodynamic guidelines for stabilizing Se-rich perovskite phases and tuning anion order to optimize band gap and phase stability in lead-free chalcogenide absorbers.
Lastly, we analyze the dielectric function as well as the absorption coefficient, and demonstrate that composition, crystal structure, and anion ordering jointly control the optical band gap.

\begin{figure*}
    \centering
    \includegraphics[scale=0.265]{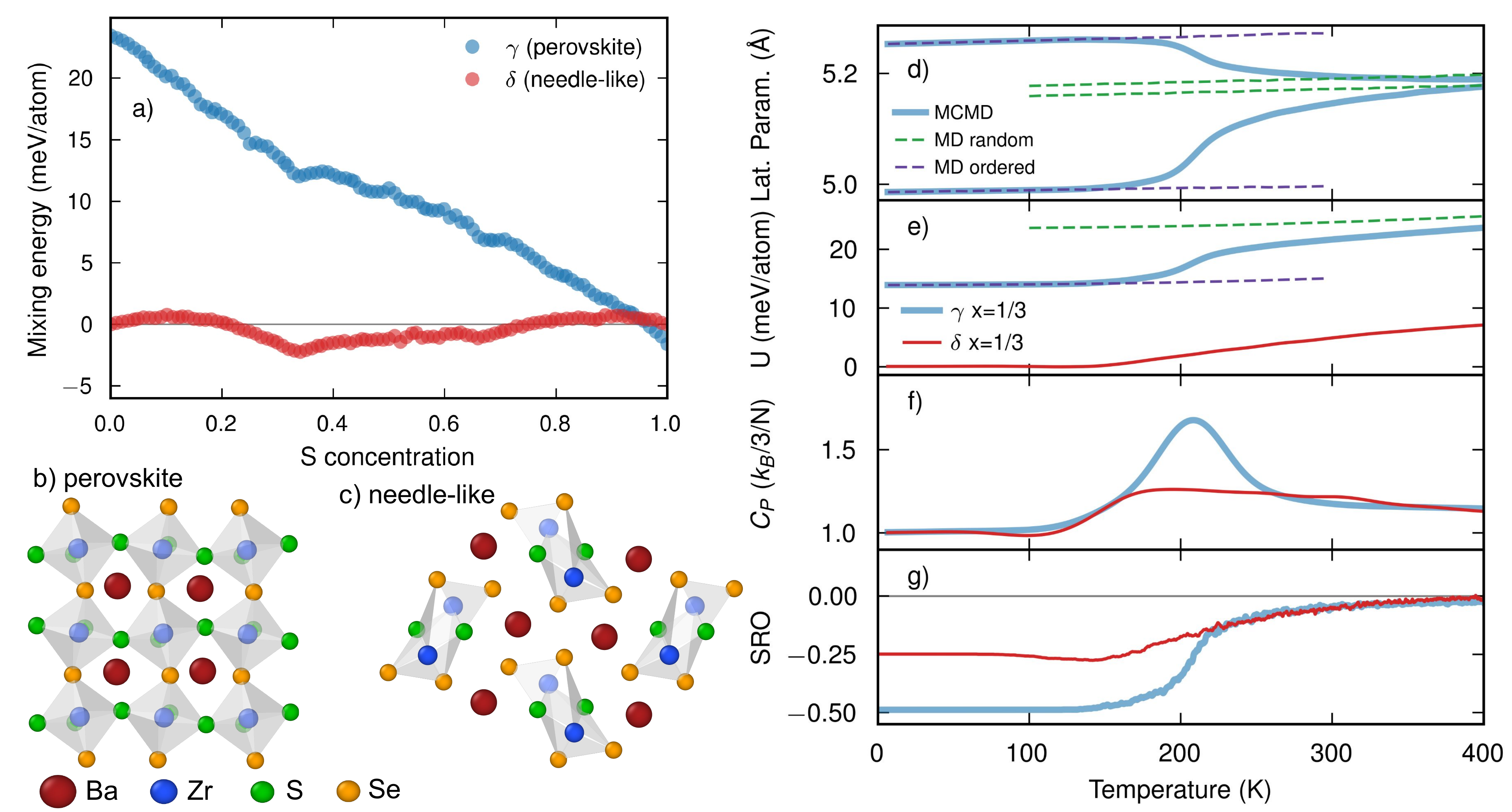}
    \caption{
        \textbf{Mixing energies and order-disorder transition from simulations.}
        (a) Comparison of the mixing energies of the perovskite ($\gamma$) and needle-like ($\delta$) phases obtained from \gls{mcmd} cooling simulations.
        The structures in (b) and (c) display the ordered structures at $x = 1/3$ with Ba atoms shown in red, Zr atoms in blue, S atoms in green, and Se atoms in orange.
        The Zr-centered coordination polyhedra are shown in gray.
        Structures were visualized using \textsc{ovito} \cite{Stukowski2010}.
        (d--g) Thermodynamic properties from heating simulations at $x = 1/3$ starting from the ordered state.
        (d) Lattice parameters referred to the primitive cubic cell for the perovskite ($\gamma$) phase using both \gls{mcmd} and \gls{md} with fully ordered or fully randomized occupations.
        (e) Change in potential energy after removal of the Dulong--Petit contribution ($\frac{3}{2}k_\text{B}T$),
        (f) heat capacity $C_P$, and (g) nearest-neighbor Warren--Cowley \gls{sro} parameter.
        }
    \label{fig:mixing-energy}
\end{figure*}

\section{Results and discussion}

\subsection{Mixing energies and ordered structures}

The \gls{mlip} model developed in this work is based on the \gls{nep} framework \cite{FanWanYin22, XuBuPan25, Larsen2017, EriFraErh19, Angqvist2019, Hart2008} (\autoref{snote:nep}) and accurately reproduces the small energy differences predicted by \gls{dft} calculations \cite{blum2009ab} using the HSE06 hybrid functional \cite{krukau2006influence} (\autoref{snote:dft-calculations}) between the perovskite, needle-like ($\delta$ \hmn{Pnma}), and hexagonal (\hmn{P6_3/mmc}) phases for the end members \ce{BaZrS3} and \ce{BaZrSe3} (\autoref{snote:ground-state-structures} and \autoref{stab:energy-differences}).
In addition, the \gls{mlip} yields mixing energies in quantitative agreement with the corresponding \gls{dft} results across the full composition range (\autoref{snote:mixing-energies}; \autoref{sfig:mixing-energies-part1} and  \autoref{sfig:mixing-energies-part2}), demonstrating the reliability of the model for thermodynamic analysis.
We note that throughout this study the mixing energy is defined relative to the pure needle-like end members (\autoref{snote:free-energies-and-phase-diagram}).

The relative \qty{0}{\kelvin} stability of the phases within the 1:1:3 stoichiometry was further assessed using simulated annealing in supercells containing several hundred atoms, which are computationally inaccessible at the \gls{dft} level (\autoref{snote:mixing-energies}).
This analysis identifies the $\delta$ needle-like phase as the lowest-energy structure across nearly the entire composition range up to approximately \qty{95}{\percent} sulfur (\autoref{fig:mixing-energy}).
These results indicate a two-phase coexistence region separating the perovskite and needle-like phases at low temperature, which we analyze in more detail below (\autoref{sect:phase-diagram}).

For both the perovskite and needle-like structures, the mixing energy exhibits a pronounced local minimum near $x = 1/3$, corresponding to layered anion ordering patterns commensurate with the respective primitive unit cells (\autoref{fig:mixing-energy}b,c).
In the perovskite case, this trans-type ordering is analogous to arrangements previously identified in mixed-halide perovskites \cite{Bechtel2018}.
However, such layered ordering appears to be uncommon in oxychalcogenide perovskites; 
oxysulfide and oxynitride perovskites typically favor cis-configurations, in which identical anions occupy adjacent sites rather than opposite vertices of the octahedron \cite{Pilania2020, Shang2025, Porter2014, Yang2011}.
Studies of oxysulfides further demonstrate that anion ordering can strongly influence electronic properties, underscoring the importance of understanding the origin of the ordered structures identified here \cite{Shang2025} (see \autoref{sect:electronic-properties}).

We propose that layered ordering is energetically favored in \bzsse{} because it enables octahedral tilting to adapt to the local chemical environment.
In the orthorhombic \hmn{Pnma} structure with tilt pattern $a^-a^-c^+$, layered ordering leads to the $c^+$ rotation acting predominantly on Se anions, while the $a^-$ tilts involve both S and Se.
Consistent with this interpretation, the relaxed Zr--Se--Zr bond angles are smaller than the corresponding Zr--S--Zr angles (\ang{155} versus \ang{156}), reflecting a slightly larger tilt amplitude associated with the larger ionic radius of Se.
Although layered ordering can induce significant interlayer strain in oxysulfide and oxynitride systems with large ionic-radius mismatch \cite{Shang2025}, the smaller size difference between S and Se likely mitigates such strain here, stabilizing the observed trans-type ordering.

\subsection{Order-disorder transitions}

To assess the finite-temperature ordering behavior of \bzsse{}, we performed heating simulations starting from the ordered ground states identified at \qty{0}{\kelvin} (\autoref{snote:md-free-energies} and \autoref{snote:mcmd}) \cite{FreAstde16, XuBuPan25, Kir35, Allan2001, SadErhStu12}.
We focus on the composition $x = 1/3$, which exhibits the strongest ordering tendency in the mixing energies.
The temperature evolution of the potential energy, heat capacity, and Warren--Cowley \gls{sro} parameter \cite{Cowley1950} reveals an order–disorder transition near \qty{210}{\kelvin} for both the $\delta$ and $\gamma$ phases (\autoref{fig:mixing-energy}d--g).  
The transition appears broad, consistent with a continuous loss of long-range order rather than a first-order discontinuity.  
Accordingly, no sharp jump in the energy is observed.

To demonstrate the impact of ordering, we additionally performed \gls{md} simulations of the perovskite phase with fully ordered and fully randomized anion occupations (\autoref{fig:mixing-energy}d,e).  
In these simulations, the system remains trapped in the imposed configurations, preventing equilibration between ordered and disordered states.  
In contrast, the combined \gls{mcmd} approach enables exchange of S and Se on the X-site sublattice and produces a smooth crossover from the ordered to the disordered regime as temperature increases.  
Above the transition temperature, the \gls{sro} parameter decreases continuously toward zero, reflecting progressive randomization of S and Se on the X-site sublattice.
Similar order–disorder behavior is observed across the composition range, but the transition is particularly pronounced at $x = 1/3$, where ideal commensurate ordering patterns are possible.

\begin{figure*}[ht]
    \centering
    \includegraphics[scale=0.24]{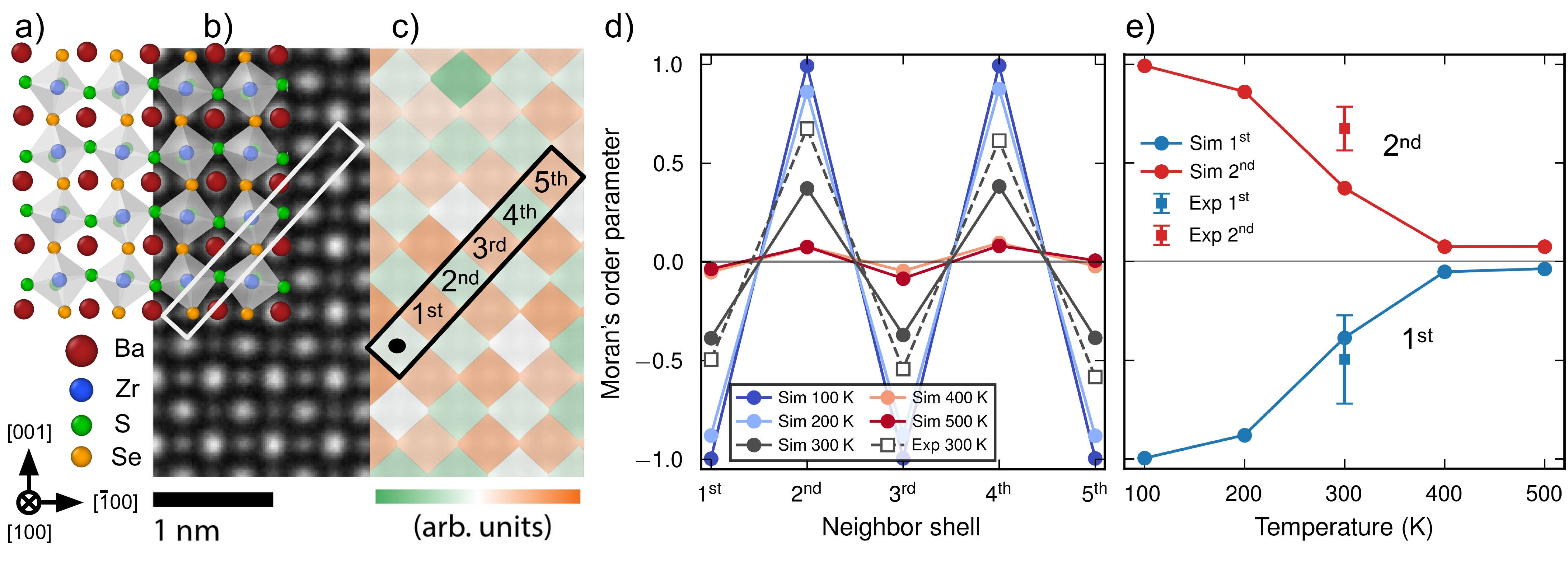}
    \caption{
    \textbf{Experimental observation of ordering.}
        (a) Atomistic structure of the ideal ordered perovskite at x=1/3.
        (b) \gls{stem} \gls{haadf} image of a sample of \bzsse{} with $x\approx0.26$ along the pseudocubic $\langle 100 \rangle$ zone axis, overlaid with the ideal atomic structure.
        (c) Anion atom column intensities corresponding to the X-sites extracted from the \gls{haadf} images (where orange and green markers scale with higher (Se) and lower (S) image intensity, respectively).
        (d) Moran's \textit{I} order parameter from experiment at approximately \qty{300}{\kelvin} (Exp) and from images based on the atomistic simulations (Sim) as a function of increasing pseudocubic $\left<110\right>$ neighbor shell.
        (e) Moran's order parameter as a function of temperature for the first and second neighbor shell.
        }
    \label{fig:exp_stemresults}
\end{figure*}

To provide experimental evidence for this unusual anion ordering and its associated order–disorder transition, we carried out \gls{stem} characterization of \bzsse{} thin-film cross sections with compositions near $x = 1/3$ (\autoref{snote:film-growth}--\ref{snote:stem-image-analysis}) \cite{ye2024processing, sadeghi_making_2021}.
Atomic-number-sensitive \gls{haadf} imaging along the pseudocubic $\langle 100 \rangle$ direction reveals an atom-column arrangement consistent with the perovskite structure of \bzsse{} (\autoref{fig:exp_stemresults}a,b).
Significant intensity variations are observed at the anion columns, with columns along projected $\langle 110 \rangle$ directions exhibiting alternating low and high intensity.
To quantify these variations, peak intensities extracted from two-dimensional Gaussian fits were mapped onto the projected structure (\autoref{fig:exp_stemresults}c), revealing spatial patterns consistent with layered anion ordering.

To enable a direct comparison between experiment and simulation, we evaluated the Moran’s $I$ nearest-neighbor autocorrelation statistic for both experimental images and simulated \gls{stem} images generated from \gls{mcmd} snapshots (\autoref{fig:exp_stemresults}d,e; see \autoref{eq:morani}) \cite{Xu2023-dl}.
As a function of increasing anion neighbor shell along $\langle 110 \rangle$ directions, the order parameters extracted from simulated images exhibit an order–disorder transition consistent with that obtained directly from the atomistic simulations.
The experimentally determined order parameters at approximately \qty{300}{\kelvin} fall between the simulated values at \num{200} and \qty{300}{\kelvin}, indicating partial ordering at room temperature.
These results confirm that the anion ordering predicted by simulations is observed experimentally at comparable temperature and composition.

It is important to note that the perovskite $\gamma$ phase is metastable with respect to the needle-like $\delta$ phase within the 1:1:3 stoichiometry, implying the presence of significant kinetic barriers between these structures.
The persistence of nonzero \gls{sro} above the nominal transition temperature, including at room temperature, suggests that local anion correlations can survive even after long-range order is lost.

\subsection{Phase diagram}
\label{sect:phase-diagram}

\begin{figure*}
    \centering
    \includegraphics[scale=0.268]{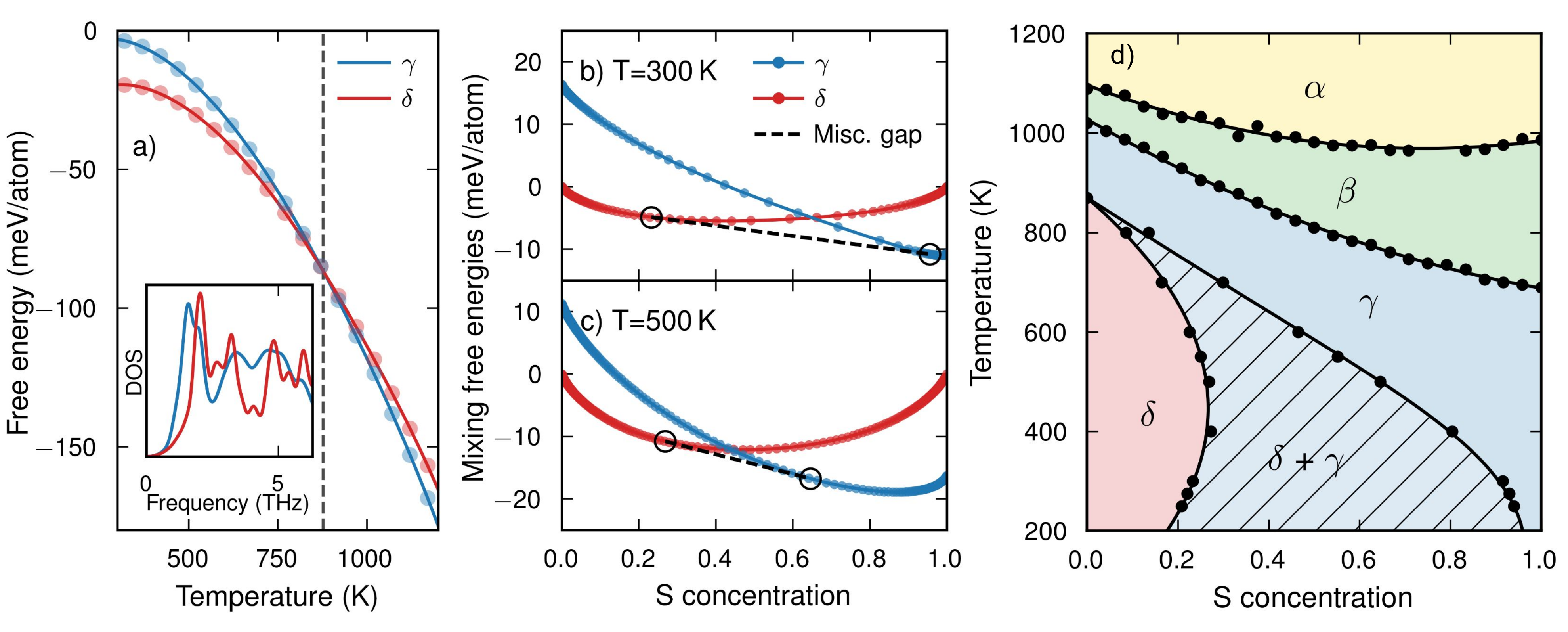}
    \caption{
        \textbf{Free energies and phase diagram.}
        (a) Free energies of the pure Se ($x$=0) $\gamma$ and $\delta$ phases.
        The inset shows the vibrational density of states obtained from \gls{md} at \qty{300}{\kelvin} via the mass-weighted velocity autocorrelation function.
        Mixing free energies for $\gamma$ and $\delta$ phases at (b) \qty{300}{\kelvin} and (c) \qty{500}{\kelvin}.
        The dashed lines indicate the miscibility gap with end points being marked with open circles.
        (c) Predicted phase diagram for the \bzsse{} system.
        Here, $\alpha$ (\hmn{Pm-3m}), $\beta$ (\hmn{I4/mcm}), and $\gamma$ (\hmn{Pnma}) refer to the three perovskite phases, while $\delta$ refers to the needle-like phase (\hmn{Pnma}).
        The striped $\delta+\gamma$ region indicates the two-phase region.
    }
    \label{fig:free-energies-phase-diagram}
\end{figure*}

We now examine the temperature–composition phase behavior of \bzsse{}.
We first determine the transitions among the perovskite polymorphs from \gls{mcmd} heating simulations (\autoref{sfig:mcmd-simulations})
which reveal successive transitions from the orthorhombic $\gamma$ (\hmn{Pnma}) to the tetragonal $\beta$ (\hmn{I4/mcm}) and finally to the cubic $\alpha$ (\hmn{Pm-3m}) phase across the composition range.
These transitions are identified through a combined analysis of lattice parameters, heat capacities, and symmetry-adapted octahedral tilt mode projections \cite{FraRosEriRahTadErh23} as functions of temperature.

In contrast, the transition between the needle-like $\delta$ phase and the perovskite $\gamma$ phase is not directly accessible from \gls{md} or \gls{mcmd} simulations because of the large nucleation barrier separating these structures.
We therefore determine this boundary from free energy calculations (\autoref{snote:mcmd}).

We begin by computing the Gibbs free energy of the pure \ce{BaZrSe3} end member, $G(x=0,T)$, using thermodynamic integration to an Einstein crystal reference (\autoref{fig:free-energies-phase-diagram}a).
The $\delta$ phase is stable up to approximately \qty{870}{\kelvin}, above which the $\gamma$ phase becomes thermodynamically favored.
Although the $\gamma$ phase is higher in energy by about \qty{24}{\milli\electronvolt\per\atom} at \qty{0}{\kelvin} (\autoref{stab:energy-differences}), it possesses a significantly larger vibrational entropy, which stabilizes it at elevated temperature.
This enhanced entropy arises from a greater density of low-frequency phonon modes in the $\gamma$ phase (\autoref{fig:free-energies-phase-diagram}a and \autoref{sfig:phonons}).

Using $G(x=0,T)$ as a reference, the mixing free energy $\Delta G_\text{mix}(x,T)$ is obtained across the full composition range via free energy integration (\autoref{snote:free-energies-and-phase-diagram}).
At \qty{300}{\kelvin}, a two-phase region is predicted between approximately $x=0.2$ and $x=0.95$, separating the $\gamma$ and $\delta$ phases (\autoref{fig:free-energies-phase-diagram}b).
At \qty{500}{\kelvin}, this coexistence region narrows and shifts asymmetrically to approximately $x=0.25$ and $x=0.65$ (\autoref{fig:free-energies-phase-diagram}c).

Free energy calculations performed over a dense grid of temperatures then yield the full temperature–composition phase diagram of \bzsse{} (\autoref{fig:free-energies-phase-diagram}d).
On the Se-rich side, the needle-like $\delta$ phase is thermodynamically stable, whereas on the S-rich side the perovskite phase is favored, with a two-phase coexistence region at intermediate compositions.
At higher temperatures, the orthorhombic $\gamma$ phase is stabilized across the entire composition range and subsequently transforms to the tetragonal $\beta$ and cubic $\alpha$ polymorphs upon further heating.

When synthesized via a conventional solid-state reaction at \qty{700}{\kelvin} \cite{Nishigaki2020}, Se substitution up to approximately \qty{40}{\percent} in the sulfide perovskite has been reported without observation of non-perovskite precipitates at room temperature.
Our calculations predict that only about \qty{9}{\percent} Se is thermodynamically soluble in the perovskite phase at \qty{300}{\kelvin}.
However, the calculated solubility limit increases to approximately \qty{20}{\percent} at \qty{400}{\kelvin} and to \qty{40}{\percent} at \qty{500}{\kelvin}, bringing the theoretical predictions closer to experimental observations under elevated-temperature growth conditions.
Furthermore, the computed free energy differences are small, and modest uncertainties in the underlying \gls{dft} reference data, the \gls{mlip} model, or the sampling can shift phase boundaries.
Indeed, first-principles predictions of phase transition temperatures in similar systems can commonly differ from experiment by \qtyrange{100}{200}{\kelvin} depending on the underlying exchange-correlation functional \cite{FraWikErh2023, Kayastha2025}.
Finally, we note that the persistence of the single-phase perovskite at room temperature indicates kinetic limitations that suppress nucleation of the thermodynamically favored $\delta$ phase at low temperature.

\subsection{Optoelectronic Properties}
\label{sect:electronic-properties}

\begin{figure*}
    \centering
    \includegraphics{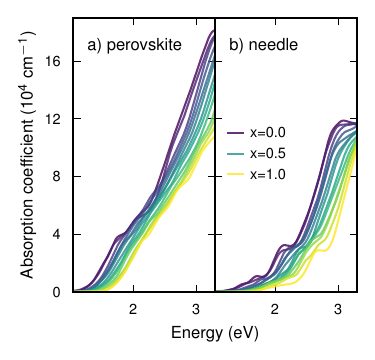}%
    \includegraphics{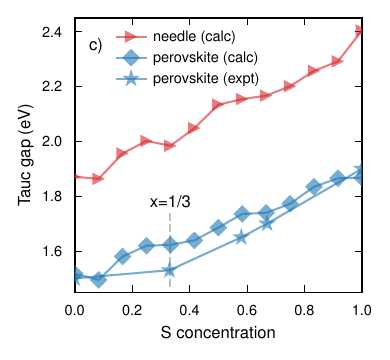}%
    \includegraphics{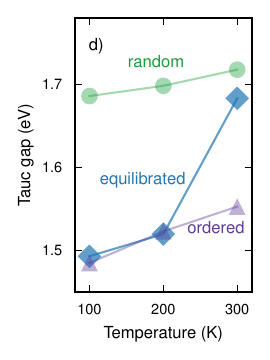}
    \caption{
        \textbf{Optical properties.}
        a, b) Absorption coefficients of the perovskite and needle-like phases, respectively, at \qty{300}{\kelvin} for varying sulfur concentrations using equilibrated \gls{mcmd} snapshots.
        c) Tauc gaps obtained from the data in (a,b) (calculated), together with experimental data from Ref.~\citenum{sadeghi2023expanding}.
        d) Tauc gaps at \qty{33}{\percent} sulfur concentration obtained from three different simulations: \gls{md} with random occupation, \gls{md} with ordered occupation, and fully equilibrated \gls{mcmd}, at 100, 200, and \qty{300}{\kelvin}.
    }
    \label{fig:dfs_taucs}
\end{figure*}

Finally, to understand how the electronic and optical properties of the material are affected by changes in composition, phase, and atomic ordering, we compute the dielectric function and the absorption coefficient.
All calculations are performed at the PBE level, followed by two corrections: a band gap correction derived from HSE06 ($\alpha = 0.25$) calculations and a \gls{soc} correction.
The latter decreases the band gap by up to \qty{0.17}{\electronvolt} for Se-containing compounds (see \autoref{snote:df_calculations} for further details).

The absorption coefficients of the perovskite and needle-like phases at \qty{300}{\kelvin} are calculated for all compositions (\autoref{fig:dfs_taucs}a,b).
For both phases, the onset of absorption shifts monotonically to higher energies with increasing sulfur concentration.
Notably, the needle phase exhibits an extended but weak absorption tail, in contrast to the sharper onset observed for the perovskite phase.
A similar difference between the two phases has previously been reported for \ce{SrZrS3} \cite{Niu2016}.

From the absorption coefficients, we extract the optical band gap via a Tauc analysis (\autoref{fig:dfs_taucs}c; see \autoref{snote:df_calculations} and \autoref{sfig:tauc-gap-fitting} for details).
For both phases, the Tauc gap increases monotonically with increasing sulfur content, in agreement with previous experimental and theoretical studies \cite{sadeghi2023expanding, Meng2016, Li2022}.
The predictions are in very good agreement with experimental data from Ref.~\citenum{sadeghi2023expanding}, particularly for the end members.

Previous \gls{dft} studies have reported that the needle phase exhibits a lower \emph{fundamental} band gap than the perovskite phase \cite{sun2015chalcogenide}.
Consistently, the band gaps obtained directly from the electronic eigenstates are smaller for the needle phase than for the perovskite phase (\autoref{stab:band-gaps}).
In contrast, the Tauc analysis shows that the \emph{optical} gap of the needle phase is notably larger.
This discrepancy arises from the extended but weak absorption tail observed in the needle phase (\autoref{fig:dfs_taucs}b), highlighting that the fundamental and optical gaps can differ significantly.

Finally, we demonstrate that the anion ordering identified in this work (\autoref{fig:exp_stemresults}) has a significant impact on the optical properties.
To this end, we perform three types of simulations at \qty{33}{\percent} S and temperatures of 100, 200, and \qty{300}{\kelvin}: \gls{md} with fully ordered anion occupation, \gls{md} with random occupation, and fully equilibrated \gls{mcmd}.
Analysis of the optical properties from \gls{md} simulations with fixed occupations (ordered or random) reveals a consistent difference of \qtyrange{0.16}{0.19}{\electronvolt} in the Tauc gap, independent of temperature.
In both cases, the Tauc gap increases with temperature, opposite to typical semiconductor behavior but consistent with trends reported for other perovskites \cite{mannino2020temperature}.
The observed difference is therefore attributable to the ordering of S and Se atoms in the system.
The fully equilibrated structures follow the Tauc gap of the ordered configuration up to \qty{200}{\kelvin}.
Between \qty{200}{\kelvin} and \qty{300}{\kelvin}, however, the Tauc gap increases rapidly toward the value of the random configuration.
This behavior is consistent with the order–disorder transition identified above, occurring between 200 and \qty{300}{\kelvin}.

\section{Conclusions}
By combining a \gls{mlip} with large-scale \gls{md} simulations, we have revealed ordering phenomena in the \bzsse{} perovskite system. 
The simulations predict the emergence of a trans-ordered configuration with alternating layers of S and Se atoms, which is particularly pronounced at $x = 1/3$. 
This ordering is supported by \gls{stem} measurements and corresponding image simulations. 
In addition, we performed free-energy calculations based on \gls{mcmd} simulations to construct the full phase diagram, considering all relevant \ce{ABX3} polymorphs. 
Our results show that while the perovskite phase is thermodynamically stable only at high sulfur content, a wide two-phase region exists, and the perovskite phase remains dynamically stable across the entire composition range. 
Finally, we demonstrate that composition, crystal phase, and anion ordering all strongly influence the optoelectronic properties of this prototypical chalcogenide perovskite. 
In particular, the order–disorder transition occurring near room temperature leads to a shift in the Tauc gap of approximately \qtyrange{0.16}{0.19}{\electronvolt}.
This demonstrates that a detailed understanding of the underlying thermodynamic properties, such as phase stability and ordering, is essential for the rational design of materials with tailored properties.

\section*{Data Availability Statement}
The \gls{nep} model and \gls{dft} data generated in this study are openly available via Zenodo at \url{https://doi.org/10.5281/zenodo.19483979}.

\section*{Acknowledgments}
This work has been supported by the Swedish Research Council (Nos. 2020-04935 and 2021-05072) as well as the Turing Scheme and the UK Engineering and Physical Sciences Research Council (EPSRC) CDT in Renewable Energy Northeast Universities (ReNU) via Grant EP/S023836/1. 
The computations were enabled by resources provided by the National Academic Infrastructure for Supercomputing in Sweden (NAISS) at C3SE, partially funded by the Swedish Research Council through grant agreement no. 2022-06725, the Berzelius resource provided by the Knut and Alice Wallenberg Foundation at NSC, and the ARCHER2 UK National Supercomputing Service through the HEC Materials Chemistry Consortium (EPSRC EP/X035859), and the UK Materials and Molecular Modelling Hub (EPSRC EP/T022213, EP/W032260, EP/P020194).
The authors acknowledge support from the National Science Foundation (NSF) under grant no. 1751736, ``CAREER: Fundamentals of Complex Chalcogenide Electronic Materials''.
A portion of this project was funded by the Skolkovo Institute of Science and Technology as part of the MITSkoltech Next Generation Program.
K.~Y. acknowledges support by the NSF Graduate Research Fellowship, grant no. 1745302. 
M.~X. and J.~M.~L acknowledge support for this work from the Air Force Office of Scientific Research (FA9550-20-0066) and the MIT Research Support Committee.
This work made use of the MIT.nano Characterization facilities.

\end{document}